\documentclass[conference,10pt]{IEEEtran}
\usepackage{epsfig,epsf,rotating,setspace,latexsym,amsmath,amssymb,amsfonts,bm,theorem,subfigure,epstopdf}
\usepackage{cite,authblk}
\usepackage{bbm}
\usepackage{algorithm}
\usepackage[noend]{algpseudocode}
\usepackage{color}
\usepackage{mathtools}
\usepackage{soul}

\algrenewcommand\algorithmicforall{\textbf{foreach}}
\algrenewcommand\algorithmicindent{.8em}

\newtheorem{theorem}{Theorem}
\newtheorem{lemma}{Lemma}

\newenvironment{Proof}[1]{\medskip\par\noindent{\bf Proof:\,}\,#1}{{\mbox{\,$\blacksquare$}\par}}

\IEEEoverridecommandlockouts
\allowdisplaybreaks

\begin{document}
 
\title{Scale-Robust Timely Asynchronous \\ Decentralized Learning}
 
\author{Purbesh Mitra \qquad Sennur Ulukus\\
        \normalsize Department of Electrical and Computer Engineering\\
        \normalsize University of Maryland, College Park, MD 20742\\
        \normalsize  \emph{pmitra@umd.edu} \qquad \emph{ulukus@umd.edu}}
\maketitle

\begin{abstract}
We consider an asynchronous decentralized learning system, which consists of a network of connected devices trying to learn a machine learning model without any centralized parameter server. The users in the network have their own local training data, which is used for learning across all the nodes in the network. The learning method consists of two processes, evolving simultaneously without any necessary synchronization. The first process is the \emph{model update,} where the users update their local model via a fixed number of stochastic gradient descent steps. The second process is \emph{model mixing,} where the users communicate with each other via randomized gossiping to exchange their models and average them to reach consensus. In this work, we investigate the staleness criteria for such a system, which is a sufficient condition for convergence of individual user models. We show that for network scaling, i.e., when the number of user devices $n$ is very large, if the gossip capacity of individual users scales as $\Omega(\log n)$, we can guarantee the convergence of user models in finite time. Furthermore, we show that the bounded staleness can only be guaranteed by any distributed opportunistic scheme by $\Omega(n)$ scaling.
\end{abstract}

\section{Introduction}\label{section: introduction}
\begin{figure}[t]
\centerline{\includegraphics[scale=0.7]{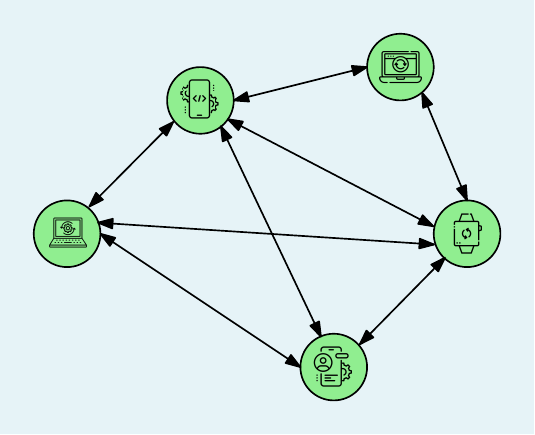}}
\caption{An example of a small distributed learning network. Each device is updating their local ML model via SGD and parallelly, asynchronously gossiping and mixing models with their neighboring devices.}
\label{fig: sample_DL_network}
\vspace*{-0.2cm}
\end{figure}

Decentralized learning, also known as gossip-based learning, is a method for learning a machine learning (ML) model with distributed data stored across different users \cite{ormandi2013gossip, hegedus2016robust}. This method utilizes two processes: model update and model mixing. The model update process is essentially the user device performing stochastic gradient descent (SGD) with the locally available data for a fixed number of steps. The model mixing process is device-to-device communication, by which the local models are exchanged and then averaged for consensus, as shown in Fig.~\ref{fig: sample_DL_network}. Decentralized learning has been analyzed in the literature as a viable alternative to federated learning \cite{hegedus2019gossip,hegedus2021decentralized}, which is the current state-of-the-art distributed ML method. With the promise of hyper-connectivity in the emergent sixth generation (6G) networks \cite{lee23_6G}, such gossip-based mechanisms provide cheap, reliable and privacy-preserving learning with decentralized implementations.

Decentralized optimization mechanism was first proved for distributed convex function optimization \cite{nedic2009distributed}. In subsequent literature \cite{nedic2010constrained, nedic2009distributed_avg, yang2019survey}, the convergence guarantee was shown with various constraints. Similar techniques were used in the decentralized learning setting for optimizing ML models \cite{ormandi2013gossip, hegedus2016robust} using SGD. The works in \cite{koloskova2019arbitrary, koloskova2019decentralized, koloskova2020unified} analyze compressed model communication for decentralized learning. The works in \cite{ram2009asynchronous, jin_scale_DDL, lian2017can, glasgow_adosd, lian2018asynchronous} showed that the process can be extended even when the asynchronous communication and model update is involved. The communication efficiency of model communication under channel delay and straggler effects in asynchronous learning setting was further improved in the subsequent studies \cite{tu2022asynchronous,wang2022asynchronous,xiong2023straggler}. The analysis in reference \cite{lian2018asynchronous} shows that linear speedup for convergence of the model can be achieved by completely asynchronous model update and model mixing process. Such asynchronous stochastic device-to-device communication process is referred to as randomized gossip algorithms. Gossip algorithms offer a low complexity mechanism to disseminate information quickly in a network \cite{shah08monograph}. Such mechanisms are often useful in low latency network applications \cite{kaswan2023age_review}, where timely information delivery and lower staleness is a sufficient criterion. In the convergence analysis in \cite{lian2018asynchronous}, the authors have assumed bounded staleness of the user models, i.e., the maximum number of training steps a user model update can lag behind the global model is bounded by a finite quantity. This assumption, however, is not immediately obvious when network scaling is considered. In large hyper-connected networks, such as in the emergent 6G communication, the network size increases and maintaining the staleness bounded, i.e., $O(1)$, requires carefully designing the communication network parameters. We call such systems, which satisfy the bounded staleness constraint, \emph{scale-robust}.

Reference \cite{kaswan2023age_review} compares a vast number of works in the literature, which analyze average information staleness (also referred to as \textit{age of gossip}) in different network settings. The most relevant works, in the context of this paper, are \cite{yates21gossip, mitra_allerton22, mitra_Infocom23, mitra2023age}. \cite{yates21gossip} shows that for a uniform gossiping scheme with a single source, the average staleness scales as $\Theta(\log n)$. \cite{mitra_allerton22, mitra_Infocom23, mitra2023age} shows that this scaling can be as low as $O(1)$ if the nodes follow an opportunistic gossiping policy. To the best of our knowledge, the only work, so far, that has considered scale-robustness for bounded staleness is \cite{mitra_hfl} in the context of asynchronous hierarchical federated learning (AHFL) setting. \cite{mitra_hfl} considers a client-edge-cloud based AHFL system, and shows that when the number of user devices grows very large, the bounded staleness criterion is achieved if the number of edge servers is $O(1)$. The AHFL setting is used for circumventing data heterogeneity, i.e., presence of non-i.i.d.~data distributions. This is achieved by clustering devices with the same data distributions together under the same edge devices. In our work, we extend the concept to decentralized learning, where multiple user devices, with heterogeneous data distributions, are in a fully connected network. We show that if the gossip capacity of the individual users scales as $\Omega(\log n)$, then the scale-robustness can be guaranteed. 

\section{System Model}\label{section: system model}
\begin{figure*}[t]
\subfigure[System model of decentralized learning.]{
\centering
\includegraphics[scale=0.7]{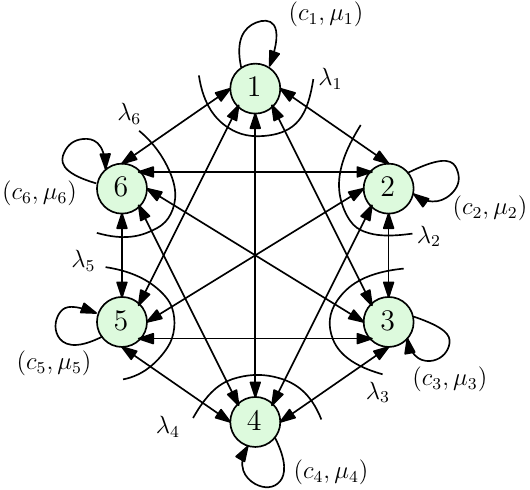}
\label{fig: DL_system_model}
}\hfill
\subfigure[Gossiping and tracking of user $1$ model.]{
\centering
\includegraphics[scale=0.85]{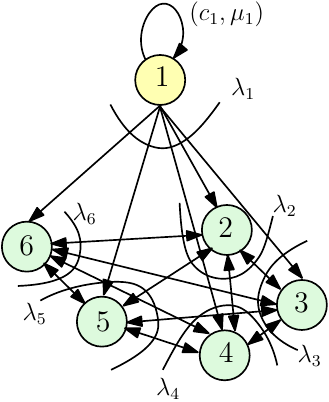}
\label{fig: internal_DL_network}
}\hfill
\subfigure[Modified gossip network.]{
\centering
\includegraphics[scale=0.85]{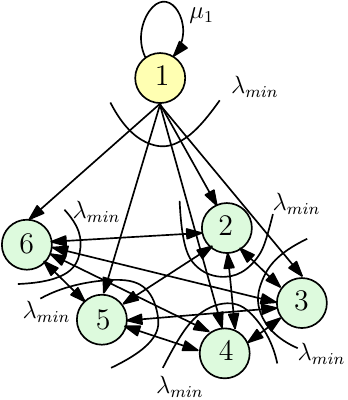}
\label{fig: modified_DL_network}
}
\caption{Different representations of a decentralized learning system.}
\vspace*{-0.2cm}
\end{figure*}

We consider a symmetric fully connected network with $n$ user devices performing decentralized learning. Each node $i$ updates its local model $\theta_i(t)$ by performing $\tau$ gradient descent steps on their locally available data $D_i$. After calculating the gradient $\nabla\mathcal{L}_i(\theta_i(t);D_i)$ and updating the model with deterministic delay $c_i$, the user device remains unavailable for a time duration. For the $i$th user, this time is a shifted exponential distribution with mean exponential time $\frac{1}{\mu_i}$. For mixing models, the gossiping from the $i$th user is characterized as an exponentially distributed unavailability time window of shifted exponential distribution with deterministic delay of $d_i$ and mean exponential time $\frac{1}{\lambda_i}$, as shown in Fig.~\ref{fig: DL_system_model}. Since in the fully connected network, each device is connected to $n-1$ neighbors, the mean unavailability window of gossiping between two devices is $d_i+\frac{n-1}{\lambda_i}\approx \frac{n-1}{\lambda_i}$, for large network size $n$. Hence, for model mixing process, we ignore $d_i$, and essentially consider a Poisson arrival process with rate $\frac{\lambda_i}{n-1}$ between a pair of users. The asynchronous decentralized learning procedure is given in Algorithm~\ref{algo: adpsgd}.

\begin{algorithm}[t]
  \caption{Asynchronous Decentralized Learning algorithm}
  \label{algo: adpsgd}
  \begin{algorithmic}[1]
    \State Initialize model $\boldsymbol{\theta}_0$ at all the users.
    \For{$i\in[n]$}
        \Procedure{ModelUpdate}{user $i$}
            \State Wait for availability time $\sim \text{Exp}(\mu_i)$.
            \State Calculate gradient $\nabla\mathcal{L}_i(\theta_i(t);D_i)$, taking time $c_i$.
            \State Update the model with learning rate $\alpha_i(t)$ as
            $$\theta_i(t+c_i)\leftarrow\theta_i(t+c_i)-\alpha_i(t)\nabla\mathcal{L}_i(\theta_i(t);D_i).$$
        \EndProcedure
        \Procedure{ModelMixing}{user $i$}
            \Procedure{ModelTransmission}{from user $i$}
                \State Wait for availability time $\sim \text{Exp}(\lambda_i)$.
                \State Randomly select a user $j$ from the set $[n]\backslash\{i\}$.
                \State Transmit model $\theta_i(t)$ to user $j$.
            \EndProcedure
            \Procedure{ModelReceival}{at user $i$}
                \State Receive model $\theta_j(t)$ from node $j\in[n]\backslash\{i\}$.
                \State Choose a random value $\beta\in(0,1)$, uniformly.
                \State Mix with the local model as
                $$\theta_i(t)\leftarrow\beta\theta_i(t)+(1-\beta)\theta_j(t).$$
            \EndProcedure
        \EndProcedure
    \EndFor
  \end{algorithmic}
\end{algorithm}

\section{Staleness Guarantee}\label{section: convergence}
In this section, we derive the gossip capacity scaling for bounded staleness guarantee. We denote the version process for the $i$th user as $N^i_i(t)$. Whenever user $i$ updates its model by SGD, $N^i_i(t)$ increases by $1$. The model version at user $j$ corresponding to user $i$'s model is denoted as $N^i_j(t)$, which is the latest version of the model of user $i$, mixed with user $j$. When user $k$ sends a gossip update to user $j$ at time $t$, the updated model becomes the latest version of the two. This is expressed as
\begin{align}
    N^i_j(t^+)=\max\{N^i_j(t^-),N^i_k(t^-)\},\quad\forall i\in[n].
\end{align}
The staleness of user $j$ at time $t$ is defined as
\begin{align}\label{def_staleness}
    S^i_j(t)=N^i_i(t)-N^i_j(t),\quad \forall i\in [n],
\end{align}
which represents the number of versions user $j$ is lagging behind the source version at user $i$. Note that $S^i_i(t)=0$. During a gossip communication from node $k$ to node $j$, the staleness, therefore, gets modified as follows
\begin{align}
    S^i_j(t^{+})=\min\{S^i_j(t^{-}),S^i_k(t^{-})\},\quad \forall i\in[n].
\end{align}
To evaluate the convergence criterion, the expected staleness in steady state needs to be finite~\cite{lian2018asynchronous}. Hence, in the context of network scaling with large $n$, we expect the following criterion for a model to converge
\begin{align}\label{bounded_staleness_criteron}
    \lim_{t\to\infty}\mathbb{E}\left[S^i_j(t)\right]=O(1),\quad \forall i\in[n], j\in[n].
\end{align}
In the following lemma, we derive an upper bound for the staleness of the decentralized learning system. 

\begin{lemma}\label{lemma: iteration}
The expected staleness of a user is bounded as 
\begin{align}\label{sum_inv_eqn}
    \lim_{t\to\infty}\mathbb{E}\left[S^i_j(t)\right]\leq\frac{\mu_i}{\lambda_{min}}\sum_{k=1}^{n-1}\frac{1}{k},
\end{align}
where $\lambda_{min}=\min\{\lambda_1,\lambda_2,\cdots,\lambda_n\}$.
\end{lemma}

\begin{Proof}
We note that even though there are multiple sources in the decentralized learning network, when a single $i$th user's model is considered, it essentially becomes a combination of $n$ different source tracking problems, as shown in Fig.~\ref{fig: internal_DL_network}. Now, since the inter-arrival times for user self-update process is a shifted exponential distribution, it does not have the memoryless property as an exponential distribution, making it difficult to analyze for staleness calculations. Hence, we calculate the staleness of the system considering exponential distribution with mean $\frac{1}{\mu_i}$ only. Since this reduces the arrival times of the new self-updates, it makes all the increments in $N_i(t)$ faster, making the resulting staleness of the system an upper bound for the original system. We denote the staleness of this new system as $\Tilde{S}^i_j(t)$. Furthermore, the different gossip rates $\{\lambda_1,\lambda_2,\cdots,\lambda_n\}$ make the network asymmetric. To derive the stated result, we replace all the rates with the minimum value $\lambda_{min}=\min\{\lambda_1,\lambda_2,\cdots,\lambda_n\}$, as shown in Fig.~\ref{fig: modified_DL_network}. Due to the substitution by lower gossip rate, this substitution yields a higher expected staleness value from the result in \cite[Thm.~2]{yates21gossip}. Thus, we obtain an upper bound for the average age for this symmetric fully connected gossip network as
\begin{align}
    \lim_{t\to\infty}\mathbb{E}\left[S^i_j(t)\right]\leq\frac{\mu_i}{\lambda_{min}}\sum_{k=1}^{n-1}\frac{1}{k}.
\end{align}
This concludes the proof of Lemma~\ref{lemma: iteration}.
\end{Proof}

Using the result of this lemma, we show a sufficient condition for gossip capacity scaling of an individual user to meet the scale-robustness condition.

\begin{theorem}\label{thm: scaling}
If the gossip capacity of individual users in a fully connected network scales as $\Omega(\log n)$, the scale-robustness condition is guaranteed.
\end{theorem}

\begin{Proof}
We can write the sum of reciprocals in \eqref{sum_inv_eqn} as
\begin{align}
    \sum_{k=1}^{n-1}\frac{1}{k}=\log(n-1)+\gamma+O\left(\frac{1}{n}\right),
\end{align}
where $\gamma$ is the Euler–Mascheroni constant. Therefore, if $\lambda_i\sim\Omega(\log n),\forall i\in[n]$, we can write
\begin{align}
    \frac{\mu_i}{\Omega(\log n)}\left(\log(n-1)+\gamma+O\left(\frac{1}{n}\right)\right)=O(1).
\end{align}
Hence, we obtain 
\begin{align}
    \lim_{t\to\infty}\mathbb{E}\left[S^i_j(t)\right]=O(1),
\end{align}
implying scale-robustness.
\end{Proof}

Now, we show that this scale-robustness cannot be achieved by any distributed opportunistic gossiping scheme by $\Omega(\log n)$ gossip rate scaling for individual users. Consider the scheme used in \cite{mitra_allerton22} for example, which allows only the freshest nodes in the network to transmit with full capacity $\sum_{i=1}^{n}\lambda_i$. Such opportunistic scheme is achieved by transmitting some pilot signal in the network whenever a user updates its model. This alerts all the other users in the network to not transmit any updates in the network, thus avoiding any possible collision or interference in the gossip capacity utilization. The freshest user keeps transmitting until it receives a signal from any other fresh user. In Theorem~\ref{thm2}, we show that this kind of scheme does not yield any scaling gain for distributed learning setting.

\begin{figure*}[t]
\subfigure[Loss vs. epochs for $\lambda_i\sim \Theta(1)$.]{
\centering
\includegraphics[scale=0.35]{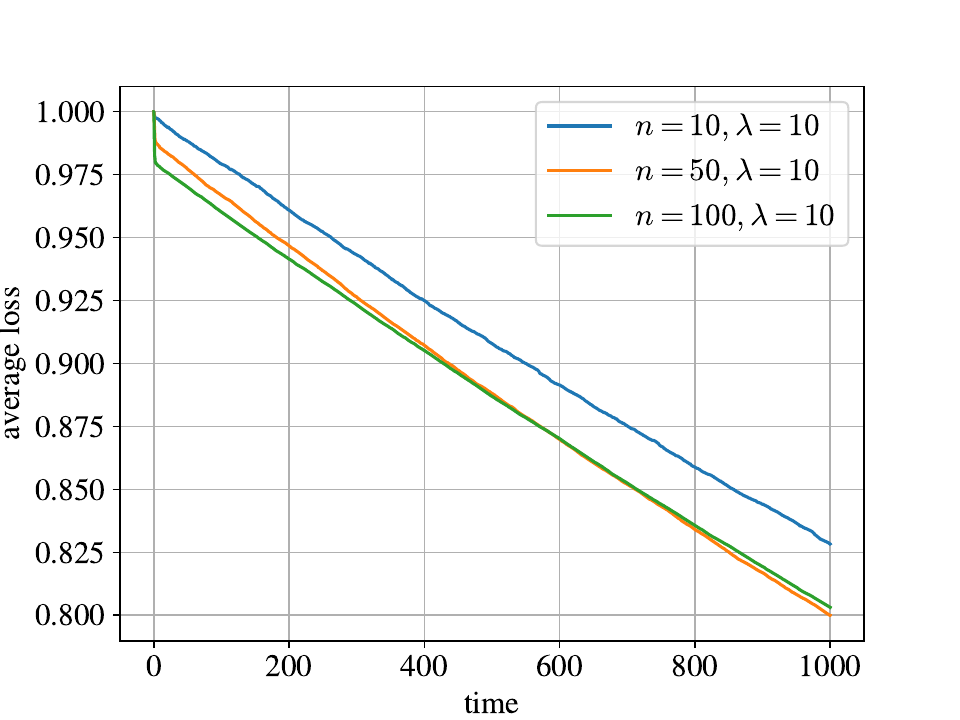}
\label{fig: lambda_1_lr}
}
\subfigure[Loss vs. epochs for $\lambda_i\sim \Theta(\log\log n)$.]{
\centering
\includegraphics[scale=0.35]{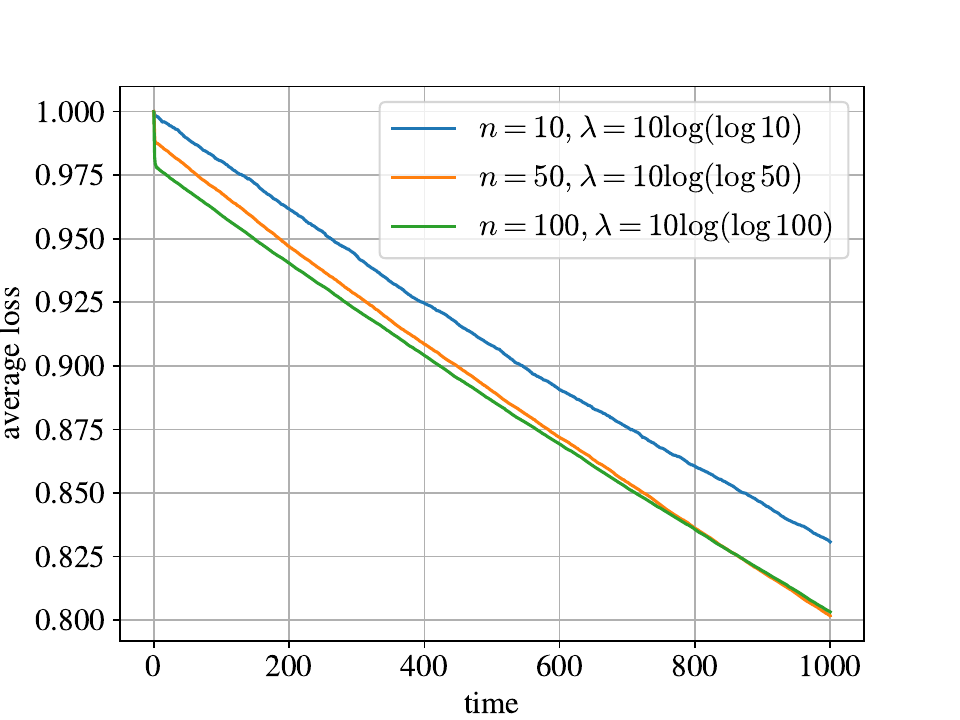}
\label{fig: lambda_log_log_n_lr}
}
\subfigure[Loss vs. epochs for $\lambda_i\sim \Theta(\log n)$.]{
\centering
\includegraphics[scale=0.35]{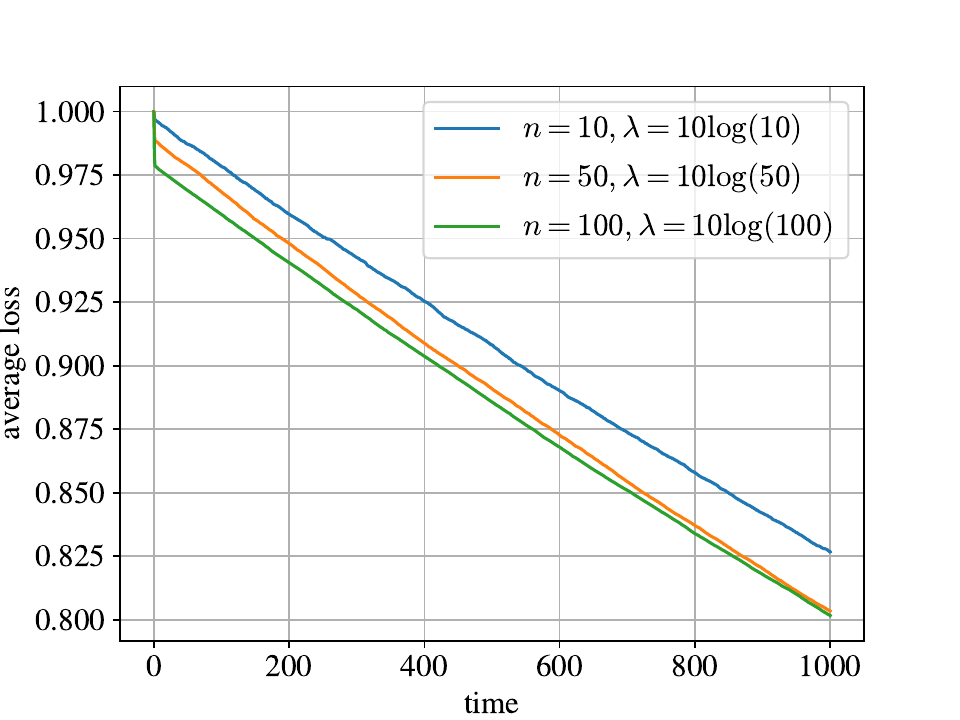}
\label{fig: lambda_log_n_lr}
}
\caption{Loss-epoch plots for linear regression task; $m=1$.}
\label{fig: loss_epoch_plots for lr}
\vspace*{-0.25cm}
\end{figure*}

\begin{theorem}\label{thm2}
The gossip capacity scaling of individual users that guarantees scale-robustness in a fully connected network using opportunistic schemes is $\Omega(n)$.
\end{theorem}

\begin{Proof}
We prove this result by evaluating the expected staleness of a user in the network, where the gossip rates are replaced by $\lambda_{max}=\max\{\lambda_1,\lambda_2,\cdots,\lambda_n\}$. Following the average age  formulation in \cite[Thm.~2]{yates21gossip}, this substitution yields a lower bound for the expected staleness. The user update rates are also replaced by the shifted exponential distribution $(c_{max}, \mu_{min})$, where $c_{max}=\max\{c_1, c_2, \cdots,c_n\}$ and $\mu_{min}=\min\{\mu_1, \mu_2, \cdots, \mu_n\}$. The mean inter update time of user $i$ is $c_{max}+\frac{1}{\mu_{min}}$. We denote the gossiping time for the $i$th user after the $k$th update as $T^{i}[k]$. As there are $n$ users in the system, the mean gossiping time is
\begin{align}\label{mean_gossip_time}
    \mathbb{E}\left[T^{i}[k]\right]=\frac{1}{n}\left(c_{max}+\frac{1}{\mu_{min}}\right).
\end{align}
We denote the staleness of this modified system at the $k$th update time as $\hat{S}^i_j[k]$. If there is no gossip communication from user $i$ to $j$, in $T^{i}[k]$, $\hat{S}^i_j[k+1]$ is just $\hat{S}^i_j[k]+1$. The probability of this event is $e^{-\lambda_{max}T^{i}[k]}$. Otherwise, with probability $1-e^{-\lambda_{max}T^{i}[k]}$, $\hat{S}^i_j[k+1]=0$. Thus, we obtain
\begin{align}\label{expected_staleness_1}
    \mathbb{E}\left[\hat{S}^i_j[k+1]\right]=\mathbb{E}\left[\left(\hat{S}^i_j[k]+1\right)e^{-\lambda_{max}T^{i}[k]}\right]. 
\end{align}
Since, $\hat{S}^i_j[k]$ and $T^{i}[k]$ are independent, (\ref{expected_staleness_1}) becomes
\begin{align}\label{expected_staleness_2}
    \mathbb{E}\left[\hat{S}^i_j[k+1]\right]=\left(\mathbb{E}\left[\hat{S}^i_j[k]\right]+1\right)\mathbb{E}\left[e^{-\lambda_{max}T^{i}[k]}\right].
\end{align}
Now, since $e^{-\lambda_{max}T^{i}[k]}$ is a convex function of $T^{i}[k]$, using Jensen's inequality and substituting from \eqref{mean_gossip_time}, we obtain
\begin{align}
    \!\!\!\mathbb{E}\left[\hat{S}^i_j[k+1]\right]&\geq\left(\mathbb{E}\left[\hat{S}^i_j[k]\right]+1\right)e^{-\lambda_{max}\mathbb{E}\left[T^{i}[k]\right]}\\
    &=\left(\mathbb{E}\left[\hat{S}^i_j[k]\right]+1\right)e^{-\frac{\lambda_{max}}{n}\left(c_{max}+\frac{1}{\mu_{min}}\right)}.
\end{align}
Thus, from the initial condition $\hat{S}^i_j[0]=0$, and using recursion
\begin{align}\label{opportunistic_gossip_scaling}
    \lim_{t\to\infty}\mathbb{E}\left[\hat{S}^i_j(t)\right]=\lim_{k\to\infty}\mathbb{E}\left[\hat{S}^i_j[k+1]\right]\geq\frac{1}{1-e^{-\frac{\lambda_{max}}{n\mu_{min}}}}.
\end{align}
Clearly, the right hand side of \eqref{opportunistic_gossip_scaling} does not yield $O(1)$ if $\lambda_{max}=\Theta(\log n)$. Only $O(n)$ scaling of $\lambda_{max}$ yields $O(1)$ lower bound for the expected staleness. Thus, the gossip rate scaling for bounded expected staleness in the original network must be $O(n)$. 
\end{Proof}

\section{Numerical Results}\label{section: numerical_results}
We show validity of the staleness bounds via numerical simulations. We consider a simple regression task that minimizes a loss function over distributed users. The loss function is 
\begin{align}\label{loss function formulation}
    \mathcal{L}(\boldsymbol{\theta};D)&=\frac{1}{|D|}\sum_{j\in[|D|]}\left(f(\boldsymbol{x}_j,\boldsymbol{\theta})-y_j\right)^2\\
    &=\frac{|D_i|}{|D|}\sum_{i\in[n]}\frac{1}{|D_i|}\sum_{j\in[|D_i|]}\left(f(\boldsymbol{x}_j,\boldsymbol{\theta})-y_j\right)^2\\
    &=\frac{|D_i|}{|D|}\sum_{i\in[n]}\mathcal{L}_i(\theta;D_i).
\end{align}
We assume that the data is equally distributed among the users, and thus, $\frac{|D_i|}{|D|}=\frac{1}{n}$. We synthetically generate the dataset as $D=\{(\boldsymbol{x}_j,y_j)\}$, where $\boldsymbol{x}_j\in\mathbb{R}^{d}$. In our case, the data points are from the Gaussian mixture distribution $\boldsymbol{x}_j\sim\frac{1}{2}\mathcal{N}\left(\frac{1.5}{d}\boldsymbol{w}_{\ell}^*,\boldsymbol{I}_d\right)+\frac{1}{2}\mathcal{N}\left(-\frac{1.5}{d}\boldsymbol{w}_{\ell}^*,\boldsymbol{I}_d\right)$, where each component of $\boldsymbol{w}_{\ell}^*\in\mathbb{R}^d$ is chosen uniformly from the interval $[0,1]$. The corresponding output is $y_j=f(\boldsymbol{x}_j,\boldsymbol{w}_{\ell}^*)$. Furthermore, $\ell\in[m]$ corresponds to the index of unique distribution present in the dataset. In our simulation $m\leq n$. Note that when only one distribution is present in the data, i.e., $m=1$, the loss function achieves its minimum at $\boldsymbol{\theta}=\boldsymbol{w}_1^*$. 

\begin{figure*}[t]
\subfigure[Loss vs. epochs for $\lambda_i\sim \Theta(1)$.]{
\centering
\includegraphics[scale=0.35]{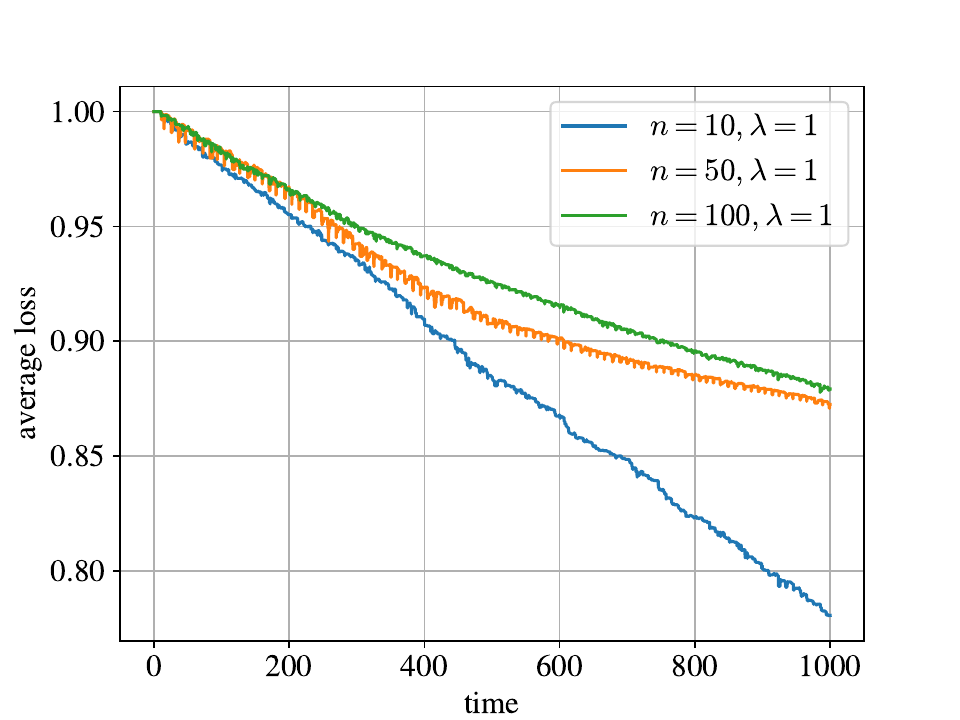}
\label{fig: lambda_1_nlr}
}
\subfigure[Loss vs. epochs for $\lambda_i\sim \Theta(\log\log n)$.]{
\centering
\includegraphics[scale=0.35]{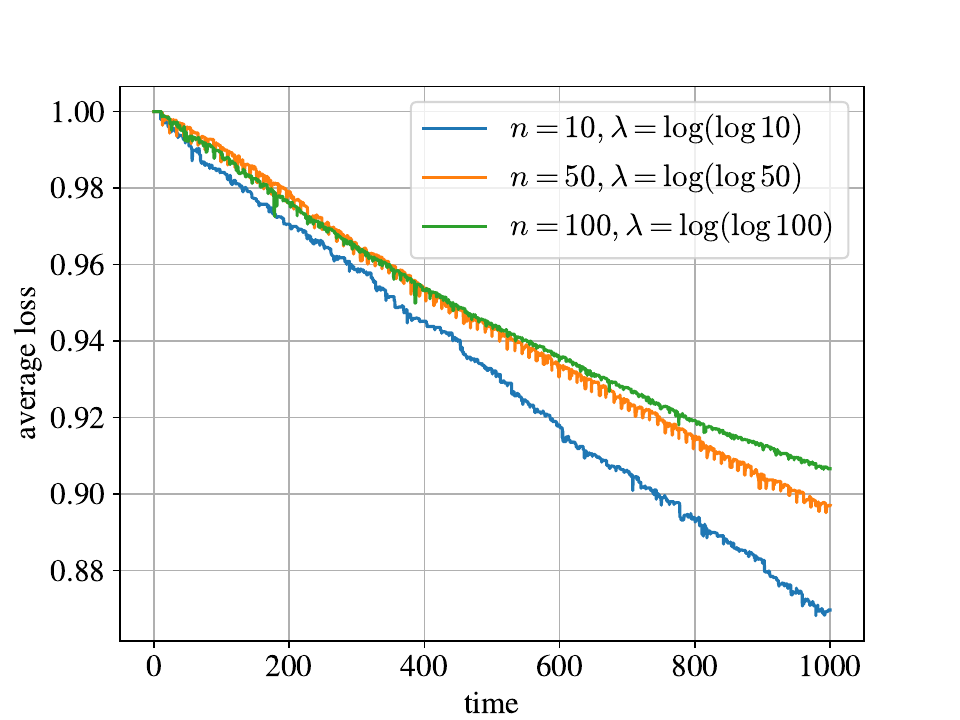}
\label{fig: lambda_log_log_n_nlr}
}
\subfigure[Loss vs. epochs for $\lambda_i\sim \Theta(\log n)$.]{
\centering
\includegraphics[scale=0.35]{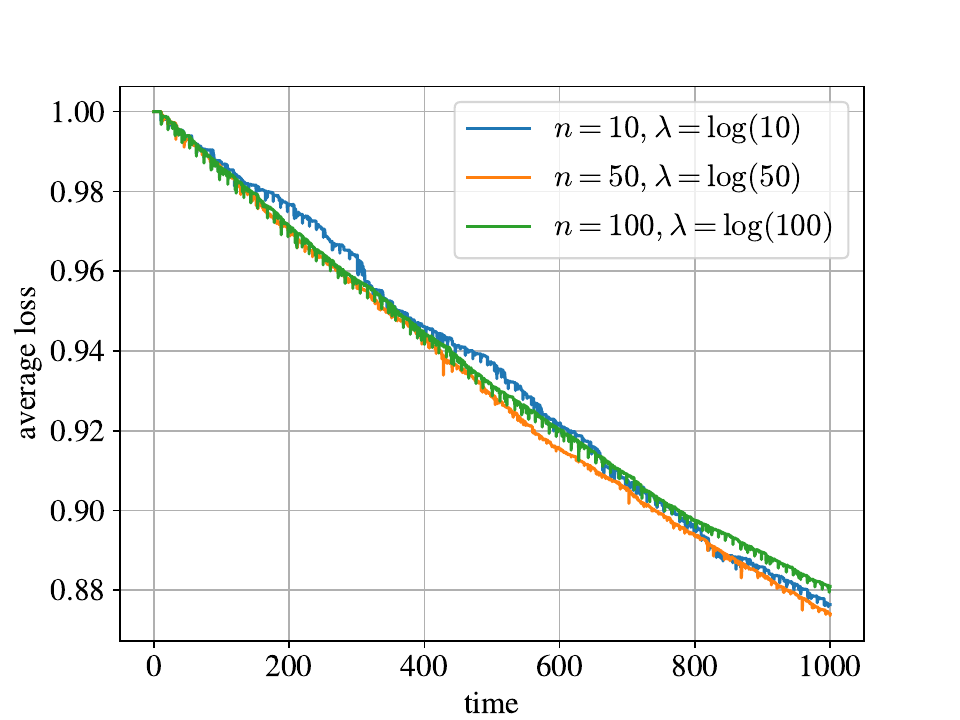}
\label{fig: lambda_log_n_nlr}
}
\caption{Loss-epoch plots for non-linear regression task; $m\sim n$.}
\label{fig: loss_epoch_plots for nlr}
\vspace*{-0.25cm}
\end{figure*}

Now, we note that this loss function satisfies the convergence criteria in \cite{lian2018asynchronous}. First, we consider a linear regression problem, i.e., $f(\boldsymbol{x},\boldsymbol{\theta})=\boldsymbol{x}^T\boldsymbol{\theta}$, with $d=100$. We observe that \eqref{loss function formulation} is differentiable, and its gradient can be written as $\nabla_{\boldsymbol{\theta}}\mathcal{L}(\boldsymbol{\theta};D)=\frac{2}{|D|}X^T(X\boldsymbol{\theta}-\boldsymbol{y})$. Since $X$ is normalized data, this shows that $\mathcal{L}$ has $L$-Lipschitz gradient. The mixing matrix $W_k$ is doubly stochastic with bounded spectral gap, by our choice of formulation in Algorithm~\ref{algo: adpsgd}. The gradient estimation is unbiased because of the addition of zero-mean mixture of Gaussian noise. The final criterion of bounded variance holds true because all the individual distributions have bounded variance, and hence the mixture of the distributions is also so. Additionally, we consider a nonlinear regression with $d=2$, where $f(\boldsymbol{x},\boldsymbol{\theta})={\theta}_1{x}_1+{\theta}_1{\theta}_2{x}_2$, where, $\boldsymbol{x}=[x_1,x_2]^T$ and $\boldsymbol{\theta}=[\theta_1,\theta_2]^T$. Since the formulation in \eqref{loss function formulation} for this case is differentiable, we obtain $\frac{\partial\mathcal{L}}{\partial\theta_1}=2({\theta}_1{x}_1+{\theta}_1{\theta}_2{x}_2-y)x_1$ and $\frac{\partial\mathcal{L}}{\partial\theta_1}=2({\theta}_1{x}_1+{\theta}_1{\theta}_2{x}_2-y)\theta_1x_1$. Following similar arguments as before, this formulation also satisfies the conditions in \cite{lian2018asynchronous}.

We simulate the linear regression task and show the loss-epoch plot in Fig.~\ref{fig: loss_epoch_plots for lr}. We show the plot for $\lambda_i\sim \Theta(1)$ scaling in Fig.~\ref{fig: lambda_1_lr}, the plot for $\lambda_i\sim \Theta(\log\log n)=o(\log n)$ in Fig.~\ref{fig: lambda_log_log_n_lr}, and the plot for $\lambda_i\sim \Theta(\log n)$ in Fig.~\ref{fig: lambda_log_n_lr}. We observe that this change in scaling of the gossip capacity does not change the loss-epoch profile of the decentralized learning setting, as in all the three cases, the loss function is deceasing with epochs and the individual models are converging, although with different rates. This can be explained by the linearity of the regression task. Since the gradient of the loss function is additive, addition of new users, and thereby increased staleness, in the system does not deviate the individual loss functions much from the overall loss functions and the users can still achieve sufficient model mixing. Hence, we observe the speedup of model convergence for large number of users, as in \cite{lian2018asynchronous}. We also observe that the linear speedup of convergence trend appears for any choice of $m$.

Next, we plot the loss-epoch profile for non-linear regression task in Fig.~\ref{fig: loss_epoch_plots for nlr}. We observe that, both for $\lambda_i\sim \Theta(1)$, and $\lambda_i\sim \Theta(\log\log n)=o(\log n)$, the loss function does not show any speedup of convergence trend for increasing number of users in Fig.~\ref{fig: lambda_1_nlr} and Fig.~\ref{fig: lambda_log_log_n_nlr}, respectively. Rather, for $n=50$ and $n=100$, it deviates from the convergence trajectory by quite a lot. However, in this setting, we observe that for $\lambda_i\sim \Theta(\log n)$, the convergence for any number of users almost follow the same trajectory in Fig.~\ref{fig: lambda_log_n_nlr}. This is consistent with the loss function, which yields non-linear gradients, thus resulting in higher deviation of the individual loss functions of the users from the global loss function.

\section{Conclusion}\label{section: conclusion}
We analyzed the scale-robustness criterion for asynchronous decentralized learning systems. In particular, we showed that for randomized gossip schemes, if the gossip capacity of the individual nodes scale as any function that is $\Omega(\log n)$, then the staleness at the users is guaranteed to be $O(1)$. Additionally, we proved that such scaling gain cannot be achieved by any opportunistic gossip scheme, as in the case of single source information dissemination. The required gossip capacity scaling for scale-robustness is $\Omega(n)$ for such opportunistic schemes. Furthermore, by numerical simulations, we observed that the necessity of scale-robustness is much more prominent with non-linear machine learning models.

\bibliographystyle{unsrt}
\bibliography{reference}

\end{document}